\documentclass[conference]{IEEEtran}

\usepackage[T1]{fontenc}
\usepackage{amsmath}
\usepackage{amssymb}
\usepackage{amsfonts}
\usepackage{amsthm}

\usepackage{graphicx}
\usepackage{multirow}
\usepackage{array}

\usepackage{tikz}
\usetikzlibrary{positioning, arrows.meta}

\usepackage{fancyvrb}
\usepackage{listings}
\usepackage{xcolor}

\usepackage{enumitem}

\usepackage{cite}
\usepackage[hidelinks]{hyperref}

\definecolor{codebg}{RGB}{248,248,248}
\definecolor{codeborder}{RGB}{190,190,190}

\lstdefinestyle{icde}{
    basicstyle=\ttfamily\footnotesize,
    keywordstyle=\bfseries\color{black},
    commentstyle=\itshape\color{black!60},
    stringstyle=\color{black},
    backgroundcolor=\color{codebg},
    frame=tb,
    rulecolor=\color{codeborder},
    framerule=0.4pt,
    xleftmargin=4pt,
    xrightmargin=4pt,
    framexleftmargin=4pt,
    framexrightmargin=4pt,
    columns=fullflexible,
    keepspaces=true,
    breaklines=true,
    breakatwhitespace=true,
    showstringspaces=false,
    numbers=none,
    captionpos=b,
    aboveskip=6pt,
    belowskip=6pt
}

\renewcommand{\arraystretch}{1.5}

\IEEEspecialpapernotice{%
This work has been submitted to the IEEE for possible
publication. Copyright may be transferred without
notice, after which this version may no longer
be accessible.%
}

\begin{document}

\title{Toward Database-Native Genomics: A Vision for Declarative Genomic Analytics}

% Add author names and affiliations here.
% The original source did not contain an active author block.
%
% Example IEEE structure:
%
% \author{
%     \IEEEauthorblockN{First Author, Second Author}
%     \IEEEauthorblockA{
%         Department of Computer Sciences\\
%         University Name\\
%         City, Country\\
%         Email addresses
%     }
% }

\author{
    \IEEEauthorblockN{
        Dhruv Aron\textsuperscript{*},
        Danial Saleem\textsuperscript{*},
        James Christopher Sorenson III,
        Xiangyao Yu
    }
    \IEEEauthorblockA{
        University of Wisconsin--Madison\\
        Madison, WI, USA\\
        \{daron2, saleem5, jcsorenson\}@wisc.edu,
        yxy@cs.wisc.edu
    }
    \IEEEauthorblockA{
        \textsuperscript{*} Equal contribution.
    }
}

\maketitle

% =========================================================
% Abstract
% =========================================================

\begin{abstract}
In recent years, numerous advancements in relational database management systems (RDBMS), including cloud-native and GPU-native paradigms, have resulted in highly competitive RDBMS performance for analytical workloads. At the same time, the field of genomics has evolved into a data-intensive domain owing to the advent of next-generation sequencing; however, current genomic analysis workflows are often file-based, fragmented, and reliant on disparate tools, limiting their expressiveness, usability, performance, and scalability. This work argues for an incremental adoption of RDBMS and declarative SQL querying in genomic analysis, particularly tertiary analysis. By translating common tasks from widely-used tools like SAMtools and BEDtools into SQL queries executed in DuckDB, we demonstrate that off-the-shelf RDBMS can match or exceed the performance of existing tools while offering significant improvements in expressiveness and usability. We identify current limitations and propose practical paths forward, aiming to position relational databases as a foundational platform for modern bioinformatics workflows.
\end{abstract}

% =========================================================
% Main Paper
% =========================================================

\section{Introduction}

The rapid adoption of next-generation sequencing has transformed genomics into an increasingly data-intensive field, producing datasets that continue to grow in both size and complexity. Despite this growth, the systems used to analyze genomic data remain largely file-based, with workflows commonly consisting of collections of independent command-line tools, libraries, and scripts, each operating on CSV files and specialized formats such as SAM/BAM, BED, VCF, and FASTQ.

Although this ecosystem has enabled substantial progress in bioinformatics, it also introduces significant complexity. Researchers must contend with incompatible or ambiguous file formats, repeated parsing and format conversion, unnecessary intermediate data, and rigid tool interfaces. They also need to manually manage concerns such as physical representation, indexing, sorting, parallelism, and pipeline composition. As genomic datasets continue to grow, these limitations increasingly affect the usability, maintainability, and performance of genomic analysis. 

We observe that this situation closely resembles the state of business data processing in the 1960s prior to the widespread adoption of relational database management systems (RDBMS), with these properties closely resembling limitations that relational databases were originally developed to overcome (see Table I). In addition to providing a logical abstraction over physical storage and a standardized declarative interface (i.e. SQL), modern RDBMS inherently provide query optimization, indexing, parallel execution, integrity constraints, access control, and a broad ecosystem of interoperable tools. Recent RDBMS research and contributions have also introduced novel paradigms and functionality such as vector, cloud-native, and GPU-native databases.

Motivated by this observation, we argue that relational database systems and declarative SQL should become foundations of modern genomic data analysis. In particular, analysis of genomic data should shift from a sequence of specialized programs and intermediate files to composable queries working on logical tables. SQL provides a common and extensible interface for ad-hoc analysis, while the underlying database engine inherently determines access paths, performs optimizations, and parallelizes workloads. We do not propose that existing genomic pipelines be replaced wholesale. Instead, we advocate an incremental transition beginning with tertiary analysis where workloads are highly query-oriented, comparatively ad-hoc, and particularly well suited to relational abstractions.

To evaluate this approach, we translate a substantial subset of operations from two widely-used genomic command-line tool suites, SAMtools and BEDtools, into standard SQL and execute them using both an off-the-shelf analytical database engine (DuckDB) and a GPU-native database (Sirius DuckDB). Our experiments show that database-backed implementations are highly competitive with traditional tools, outperforming their command-line counterparts on most evaluated operations. At the same time, our results expose limitations and future extensions regarding interval operations, database storage utilization, and incompatibility with specific genomic operators and workloads. These results suggest that modern RDBMS already provide a viable foundation for a significant portion of tertiary genomic analysis, while also identifying the database capabilities that must evolve to truly make genomics database-native.

\begin{table*}[t]
\centering
\caption{Comparison of Data Processing Paradigms in Genomics and Databases.}
\label{tab:genomics-rdbms-comparison}
\small
\renewcommand{\arraystretch}{1.4}
\begin{tabular}{|>{\bfseries\centering\arraybackslash}p{2.2cm}|p{4.2cm}|p{4.2cm}|p{4.2cm}|}
\hline
 & \textbf{Genomics Today} & \textbf{Pre-Relational DBMS} & \textbf{Modern Relational DBMS} \\
\hline

Data Representation
& Flat files, with new tools often inventing new formats (e.g. SAM, BAM, BED, VCF, FASTQ)
& Flat files (hierarchical \& network file stores), each formatted per application
& Logical tables with standardized relational interfaces, explicit schemas, and integrity constraints \\
\hline

Abstraction Model
& Physical-Logical dependence, with tools depending directly on physical layout (e.g. sort order, index naming, binary format)
& Physical-Logical dependence, with programs navigating disk layout explicitly
& Physical-Logical independence, with the schema decoupled from storage (can change underlying formats without breaking programs) \\
\hline

Tool Interface
& Command-line binaries (e.g. SAMtools, BEDtools, BCFtools) with fixed functionality
& Custom procedural programs hardwired to physical data layout
& Declarative, composable, and standardized SQL \\
\hline

Performance Strategy
& Hand-tuned C/C++, with researchers manually recreating parallelization, indexing, and sorting
& Hand-written programs, with developers writing custom access paths without an optimizer
& Query optimizer that automatically performs index selection, join reordering, parallelism, and more \\
\hline

Pipeline Composition
& Orchestration tools (e.g. Makefile, Snakemake, or Nextflow) that stitch binaries and programs together via intermediate files
& Ad-hoc shell scripts that chained programs together
& Engine computes execution plan, with SQL and CTEs composing relational processing stages; external workflow orchestration remains available\\
\hline

Ecosystem \& Extensibility
& Isolated, as each tool is a standalone binary with no extension points; new capabilities require new tools and new file formats
& Isolated, as each program was built for one purpose without any shared foundation for others to build on
& Composable platform, with the same SQL interface extended into cloud warehouses, GPU engines, vector search, ML pipelines, etc. \\
\hline

\end{tabular}
\end{table*}
\section{Current State of Genomic Pipelines}

%(todo: this needs to be made more concise). 
%(todo: find an example for TB scale pipeline, maybe from broad institute).

Bioinformatics has seen exponential growth since the invention of next generation sequencing~\cite{ngs}, which has led to an explosion in the size and heterogeneity of genomic datasets ~\cite{sequencing.cost}. Consequently, genomic workflows have also become more sophisticated and complex. This necessitates more efficient and scalable approaches to genomic data analytics, and we identify that there are significant opportunities for performance and efficiency improvements.

\subsection{Example Multi-Variant Workflow}

\begin{figure}[!t]
  \centering
  \includegraphics[width=\columnwidth]{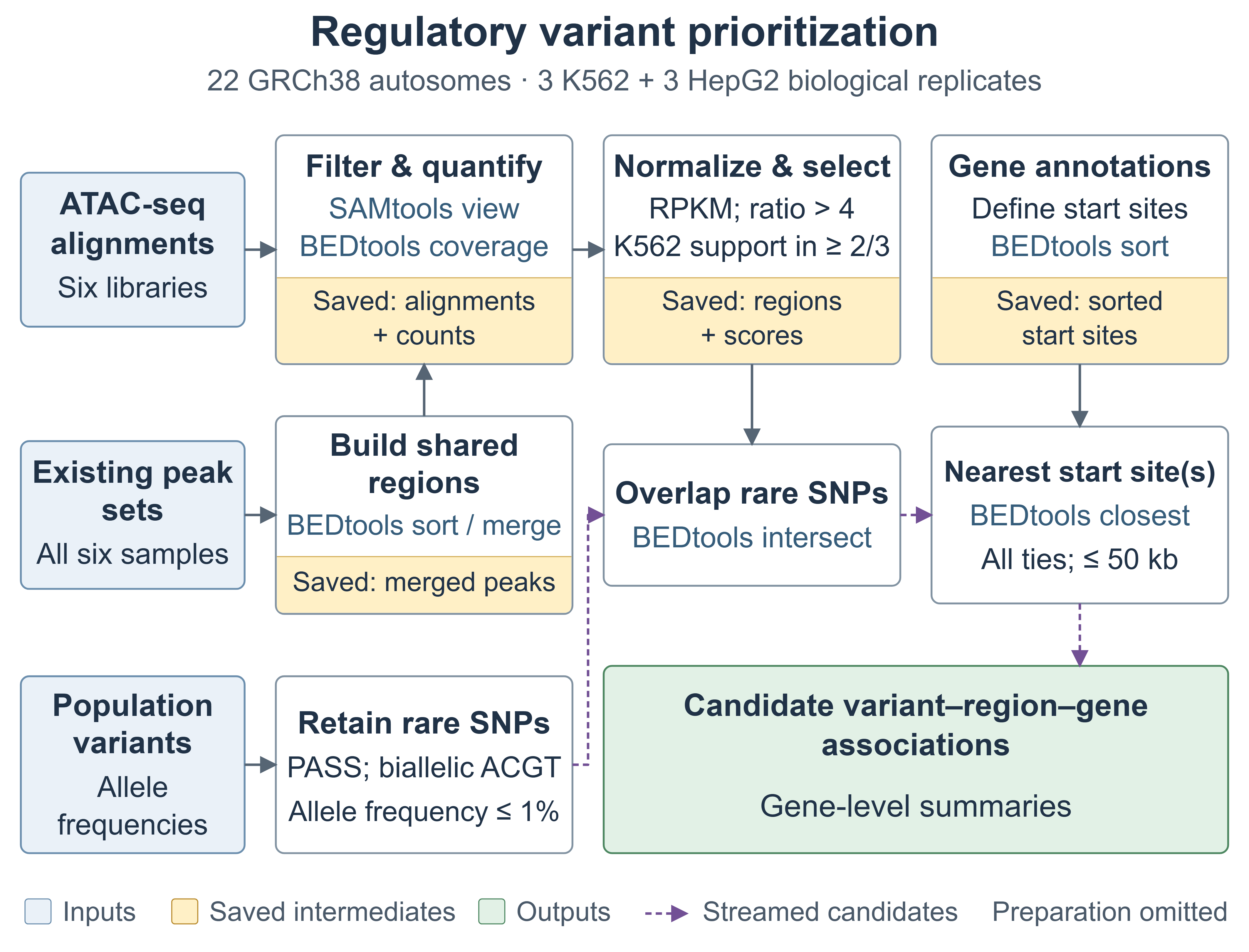}
  \caption{Multi-variant workflow integrating sequencing
alignments, accessible genomic regions, population variants,
and gene annotations to identify candidate regulatory
variant--gene associations.}
  \label{fig:multi-variant-workflow}
\end{figure}

To motivate our approach, we briefly describe how genomic data analytics is performed today using a regulatory variant prioritization multi-variant workflow as an example (based loosely on \cite{Fu2014} and \cite{Corces2018}, see fig. \ref{fig:multi-variant-workflow}).

The objective of this workflow is to identify rare genetic variants located in regions that are substantially more accessible in one cell type than another by comparing two human cell lines, K562 and HepG2, using three biological replicates of each. The workflow begins by using BEDtools to sort and merge overlapping peaks from all six samples, producing a common set of genomic regions for comparison. Next, SAMtools filters the sequencing alignments to remove low-quality or otherwise unsuitable records, after which BEDtools counts the number of alignments overlapping each region in each sample. These counts are normalized to account for differences in sequencing depth and region length. The workflow then identifies regions whose normalized accessibility signal is more than four times greater in K562 than in HepG2, requiring sufficient support from at least two of the three K562 replicates. This produces a set of regions exhibiting consistently stronger accessibility in the target cell line.

Separately, a population variant dataset is filtered to retain high-quality single-nucleotide variants with an allele frequency of at most 1\%. BEDtools then intersects these variants with the selected regions, retaining only variants that fall within regions exhibiting stronger K562 accessibility. Finally, gene annotations are used to identify the nearest gene transcription start sites (the positions where genes begin to be transcribed) within 50,000 DNA bases of each retained variant. The output consists of candidate variant--region--gene associations and gene-level summaries that researchers can use to prioritize further investigation. These associations represent potential regulatory relationships, rather than experimentally confirmed effects.

\subsection{Analysis of Status Quo}

As illustrated by the example workflow, modern genomics data analytics has the following two key properties: 

\textbf{File-based science:} Genomic data processing evolved to utilize files as primary data stores due to their immediate utility, namely the ease in modifying, transferring, compressing, and using them as input/output for disparate tools. As a result, there is no universal genomic data format nor a move from flat-file based storage to more richly structured and sophisticated storage methods. Thus, genomics remains primarily a file-based science. 
% \yxy{say a bit more here. File-based nature is the root cause of most of the problems in this area, and deserves more discussion.}

% (todo: can we find a paper on this) for being file based science.

\iffalse
Some common file formats are presented in the following table for reference. The list is not exhaustive. \\
\begin{table}[H]
  \centering
  \caption{Examples of Common File Formats used in Genomics}
  \label{tab:genomic_formats}
  \begin{tabular}{|l|l|}
    \toprule
    \textbf{Type of Data} & \textbf{Common Formats} \\ 
    \midrule
    Sequences & FASTA/FASTQ          \\ \hline
    Alignment & SAM, BAM, CRAM   \\ \hline 
    Variants  & BCF              \\ \hline
    Generic Feature & GFF, GTF \\ \bottomrule
  \end{tabular}
\end{table}
\fi

\textbf{Fragmented workflows:} 
Genomic analytics workflows are fragmented, consisting of independent, modular tools that each utilize files for input and output instead of a single standardized platform that does not need to fabricate intermediate files. While not a standard categorization, these tools can be roughly divided according to their usage, level of standardization, and form of distribution: 

% (todo: look at bioconvert) 
\begin{itemize}
    \item Standardized tools that are distributed as binaries and widely utilized (e.g. BEDtools)~\cite{bedtools}. 
    \item Libraries in different languages that provide abstract interfaces to genomic data and are used to develop new standalone tools or in one-off scripts used in the tertiary stage of workflows (e.g., Biopython\cite{biopython} and htslib\cite{Bonfield2021}).
    \item Orchestration tools such as Snakemake \cite{snakemake}, Nextflow \cite{nextflow} and Common Workflow Language \cite{cwl} that stitch together multiple tools and scripts to form a pipeline.
    \item GUI based interfaces (e.g. Galaxy \cite{galaxy.orig}) that make a subset of standard tools more accessible to those with less programming experience. 
% (todo: is GATK also a similar library? in the end it is also a collection of tools). 
\end{itemize} 

% The tooling fragmentation introduces overhead in data engineering, and makes workflows hard to extend, optimize, and scale as analytical complexity grows.

\subsection{Problems with the Status Quo}

Using flat files as data sources and relying on independent, fragmented tools to compose complex pipelines results in numerous interrelated issues. We summarize the issues in Table~\ref{tab:genomics-rdbms-comparison} and discuss them below. \\% such as the ones discussed below. 

\noindent\textbf{Challenge \#1: Lack of Data Standardization.}

\textit{Syntactic Ambiguity.} Various genomic data formats do not have concrete guidelines, resulting in tools enforcing arbitrary and oft contradictory syntax rules. For example, FASTA files permit sequences to span multiple lines without requiring a fixed line length. However, SAMtools' faidx indexing utility requires all sequence lines within a given record to have the same length, except for the final line. Consequently, FASTA files accepted by some tools may require reformatting before they can be indexed and queried using SAMtools.  

\textit{Semantic Ambiguity.} Certain data types have multiple encoding schemes without a way to identify which one is being used. For example, Phred quality scores in FASTQ files can be encoded in two ways, Phred+33 and Phred+64 ~\cite{Cock2009}, with the ASCII character `A' representing a score of 32 in the former scheme and a score of 1 in the latter. Yet, FASTQ does not explicitly provide encoding scheme identification methods, making it difficult to determine what the data actually represents. 

\textit{File Format Overhead.} In some cases, multiple formats work for the same task (e.g. both FASTA/Q and BAM/uBAM are adopted for sequencing data). However, tools that utilize one format/version may not support the other(s), forcing redundant format conversions and thus leading to unnecessary complexity and performance overheads \cite{Bioconvert, ConvertTools}. 
    
\noindent\textbf{Challenge \#2: Performance Scaling.}

\textit{Not Optimized for Performance.} Data size increases necessitate modern computing paradigms such as parallelization and distributed systems, while genomic data formats often focus on human-centric design. They are not inherently parallelized or distributed and often sacrifice other operations that are essential for ideal performance as data size grows, forcing scientists to manually implement such paradigms or optimizations. Alternatively, recent research investigates genomics analytics on Apache Spark clusters (see Related Work), though this requires significant changes to the single-node data and tools stack that is currently widely utilized. We demonstrate later in this work how certain operations based on these file formats could potentially achieve more optimal performance. 

\textit{Excessive Resource Usage.} The use of independent tools results in the side-effect of many intermediate files being created, stressing storage requirements \cite{GenAP}. Utilizing files for input and output also limits opportunities for optimization across processing stages and introduces additional disk input and output overheads. \\ %(todo: should we cite the disk io point with the gesall paper. should we find examples/citations for the second point.) 

\noindent\textbf{Challenge \#3: Difficult User and Developer Experiences.}

\textit{Lack of Expressiveness.} Existing tools often have unintuitive or ambiguous interfaces. For example, with SAMtools, many operations require the data to be preemptively indexed, filtered, and sorted prior to execution. This forces excessive documentation consultation and hampers effective debugging \cite{BamQL, SamQL}. 
    
\textit{Lack of Abstraction and Composability.} Publishing new tools as binaries forces such programs to re-implement file parsing and other fundamental operations that should be abstracted away. Moreover, these tools are not readily modifiable or extendable, necessitating forks or new tools to extend functionality and external orchestration tools (e.g. Snakemake, Nextflow) to pipe analysis steps together. Developers must thus learn other technologies and concepts outside of their primary interests and expertise, introducing incidental complexity. 
\section{Bioinformatics on RDBMS: A Vision}

Genomic analysis should evolve from brittle, inexpressive, and procedural file-based pipelines to an integrated, interactive, and scalable platform. Considering the current state of genomic data analysis' similarities to business data analytics and the success of RDBMS for the latter, bioinformatics can directly build on modern database engines. We now discuss how this approach addresses the previously enumerated issues.

\textit{Logical-Physical Independence and Expressiveness.} Logical tables hold all data, inherently eliminating parsing complexities and enforcing integrity via constraints and schemas (e.g. enforcing valid genomic coordinates and reference identifiers). As a result, SQL offers a much more standardized, readable, and maintainable format for data management and querying and additionally alleviates the need to learn or orchestrate multiple tools. We present the same query written in SAMtools (see listing \ref{lst:samtools-filter}) and in SQL (see listing \ref{lst:sql-filter}) as an example, demonstrating the better readability of the latter. Database engines also automatically maintain binary/compressed variants and textual representations, with data transformations represented as SQL fragments computed on demand instead of multiple files containing the same data in different shapes or sorted orderings.
% Older version of paper mentioned SAMQL in the paragraph above, unsure if it should be mentioned or not

\begin{lstlisting}[
    style=icde,
    language=bash,
    float=tbp,
    caption={Filtering alignments using SAMtools and AWK.},
    label={lst:samtools-filter}
]
samtools view input.bam chr1:1000001- \
  | awk '($4 > 1000000 &&
          $6 ~ /^15M/ &&
          (length($10) < 75 || $10 ~ /^ATG/))'
\end{lstlisting}

\begin{lstlisting}[
    style=icde,
    language=SQL,
    float=tbp,
    caption={Equivalent filtering using SQL.},
    label={lst:sql-filter}
]
SELECT *
FROM alignments
WHERE RNAME = 'chr1'
  AND POS > 1000000
  AND CIGAR LIKE '15M%'
  AND (
      LEN(SEQ) < 75
      OR SEQ LIKE 'ATG%'
  );
\end{lstlisting}

\textit{Longstanding Research and Development.} Building on existing relational databases makes use of decades of research in performance optimization, security, and governance and inherently provides access to thriving ecosystems of auxiliary tools. In particular, databases offer better performance through automatic query optimization, often inherently support paradigms such as parallelization, and support role-based access control to offer fine-grained access to data and centralize data governance. Moreover, all major programming languages offer integrations with all major database systems, and recent RDBMS research has led to numerous novel applications for the technology such as incremental view maintenance, cloud-native and GPU-native databases, vector databases, and built-in machine-learning pipelines. Current genomics workflows cannot easily integrate or utilize these advancements. 
%good example would be the PRQL project, which provides language that builds on top of SQL to run more complicated pipelines atop RDBMS. 
\section{Experiments and Analysis}
As a proof of concept, we convert a significant subset of operations provided by two of the most widely used tools in bioinformatics, SAMtools~\cite{SAMOriginal} and BEDtools~\cite{bedtools}, to equivalent SQL queries. Equivalence for all experiments was tested through exact output matching in the case of deterministic outputs and through multi-set matching in the case of different ordering. We focus primarily on operations that are used in ad-hoc querying and can be readily represented in SQL, leaving more complex algorithms for future work. 

Details of SAM/BAM and the BED file formats are present in Table \ref{tab:sam_columns} and Table \ref{tab:bed_columns}. In the case of BED files, all of these columns need not be present. Most tools only require the first three, and, in the dataset that we use, only the first six columns are present. 

\begin{table}[h!]
\centering
\caption{Mandatory Columns in SAM File Format}
\label{tab:sam_columns}
\footnotesize
\begin{tabular}{|c|l|l|}
\hline
\textbf{Col} & \textbf{Field} & \textbf{Description} \\
\hline
1  & QNAME  & Query (read) name \\
2  & FLAG   & Bitwise flag (read properties) \\
3  & RNAME  & Reference sequence name \\
4  & POS    & 1-based leftmost mapping position \\
5  & MAPQ   & Mapping quality (Phred-scaled) \\
6  & CIGAR  & CIGAR string (alignment) \\
7  & RNEXT  & Ref. name of the mate/next read \\
8  & PNEXT  & Position of the mate/next read \\
9  & TLEN   & Observed template length \\
10 & SEQ    & Segment sequence \\
11 & QUAL   & ASCII of Phred-scaled base quality+33 \\
\hline
\end{tabular}
\end{table}

\begin{table}[h!]
\centering
\caption{Standard Columns in BED File Format}
\label{tab:bed_columns}
\footnotesize
\begin{tabular}{|c|l|p{0.55\columnwidth}|}
\hline
\textbf{Col} & \textbf{Field} & \textbf{Description} \\
\hline
1  & chrom        & Name of the chromosome (e.g. chrX) \\
2  & chromStart   & Start position of the feature (0-based) \\
3  & chromEnd     & End position of the feature (exclusive) \\
4  & name         & Name of the feature (optional) \\
5  & score        & Score from 0 to 1000 (optional) \\
6  & strand       & Strand: + for forward, - for reverse (optional) \\
7  & thickStart   & Start position where feature is thick (e.g., start
of coding region; optional) \\
8  & thickEnd     & End position where feature is thick (optional) \\
9  & itemRgb      & RGB color value for display (e.g., 255,0,0;
optional) \\
10 & blockCount   & Number of blocks (e.g., exons; optional) \\
11 & blockSizes   & Comma-separated list of block sizes (optional) \\
12 & blockStarts  & Comma-separated list of block starts relative
to chromStart (optional) \\
\hline
\end{tabular}
\end{table}

\subsection{Experimental Setup}
For our database, we use version 1.5.6 of DuckDB (eight threads and 40GB memory limit), motivated by its ease of use, community support, and support for analytical workloads due to its columnar storage design. We also utilize Sirius \cite{Yogatama2025} (20 GiB GPU pool, 48 GiB host pool, four GPU pipeline workers, 512 MiB scans, 256 MiB hash partitions/concatenation, and 512 MiB build/sort limits), a GPU-native extension for DuckDB, to demonstrate the potential of GPU-native databases. Much data in genomics is already tabular in structure (such as the .sam and .bam files utilized by SAMtools and BEDtools commands) and is therefore easily translated into SQL tables. We show an example of how .bed data can be loaded into a DuckDB table (see listing \ref{lst:load-bed}). 

\begin{lstlisting}[
    breaklines=true, 
    language=SQL,
    float=tbp,
    caption={Example of loading .bed file into DuckDB table.},
    label={lst:load-bed}
]
CREATE TABLE bed (
    chrom      VARCHAR,
    chromStart INTEGER,
    chromEnd   INTEGER,
    name       VARCHAR,
    score      INTEGER,
    strand     VARCHAR
);

INSERT INTO bed
SELECT column0, column1, column2, column3, column4, column5
FROM read_csv('processed.bed', delim='\t', quote='', header=false, auto_detect=false,
    columns={
        'column0': 'VARCHAR', 'column1': 'INTEGER', 'column2': 'INTEGER',
        'column3': 'VARCHAR', 'column4': 'INTEGER', 'column5': 'VARCHAR'
    });
\end{lstlisting}

For our experiments, we utilize an Ubuntu 24.04 LTS image with two AMD EPYC 7302 16-core CPUs at 3.0 GHz, 128 GB ECC RAM (eight 16GB DDR4-3200 RDIMMs), storage composed of two 480 GB SATA SSDs and a 1.6 TB PCIe 4.0 ×4 NVMe SSD, and a 24GB Nvidia A30 (Ampere) GPU. The relevant SQL queries, command-line implementations, benchmarking scripts, and instructions for reproducing our experiments are publicly available at \url{github.com/dhruv-aron/toward-database-native-genomics}.  

\subsection{Experiment 1: Individual Operation Conversions}

We converted numerous SAMtools and BEDtools commands into optimized GenomicSQLite, DuckDB SQL, and Sirius-compatible queries and benchmarked individual execution times (with a timeout set at thirty minutes) on the same underlying data. The operating system page cache was prewarmed appropriately before every benchmark run. We used a subset of the SamQL \cite{SamQL} dataset containing approximately 22M reads (rows) and converted appropriately to .sam or .bed as necessary. For operations that require multiple files, we split this 22M-read subset into two separate halves (randomly deciding which read is placed in which half) and then utilized these two halves for the corresponding operations.

\begin{figure*}[!t]
  \centering
  \includegraphics[width=0.95\textwidth]{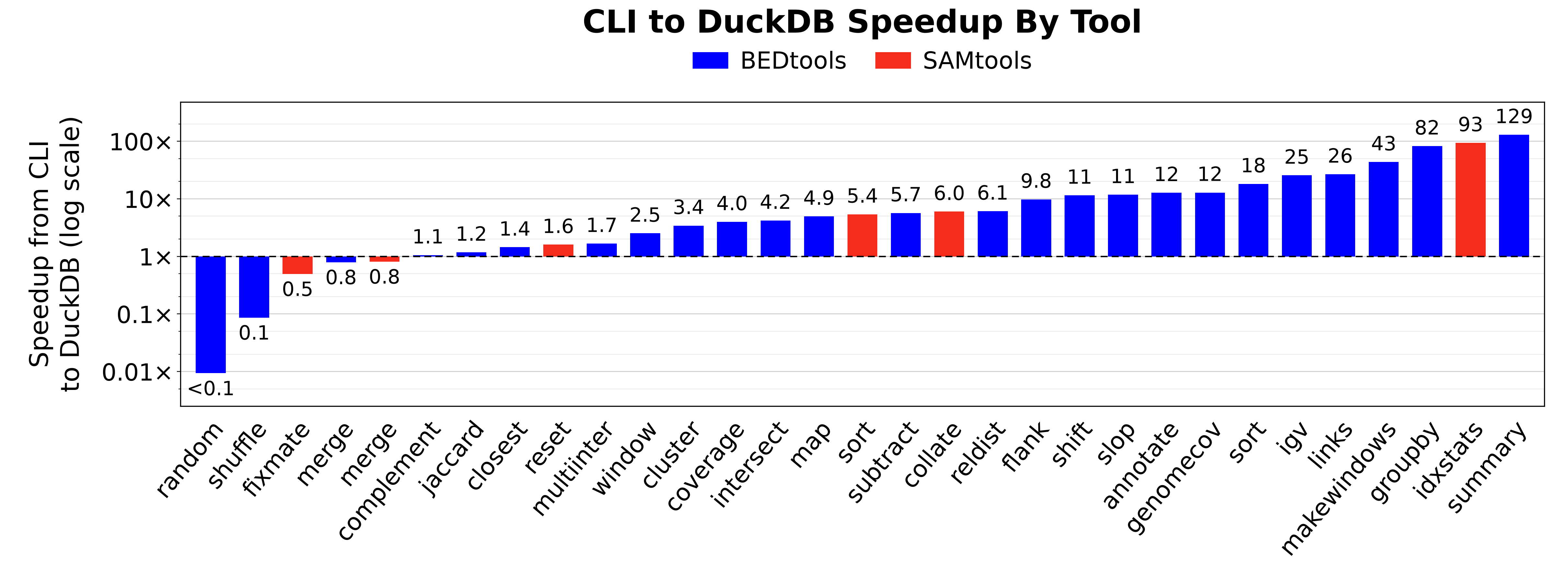}
  \caption{CLI to DuckDB speedups by tool.}
  \label{fig:cli_to_duckdb_speedups}
\end{figure*}

\begin{figure*}[!t]
  \centering
  \includegraphics[width=0.95\textwidth]{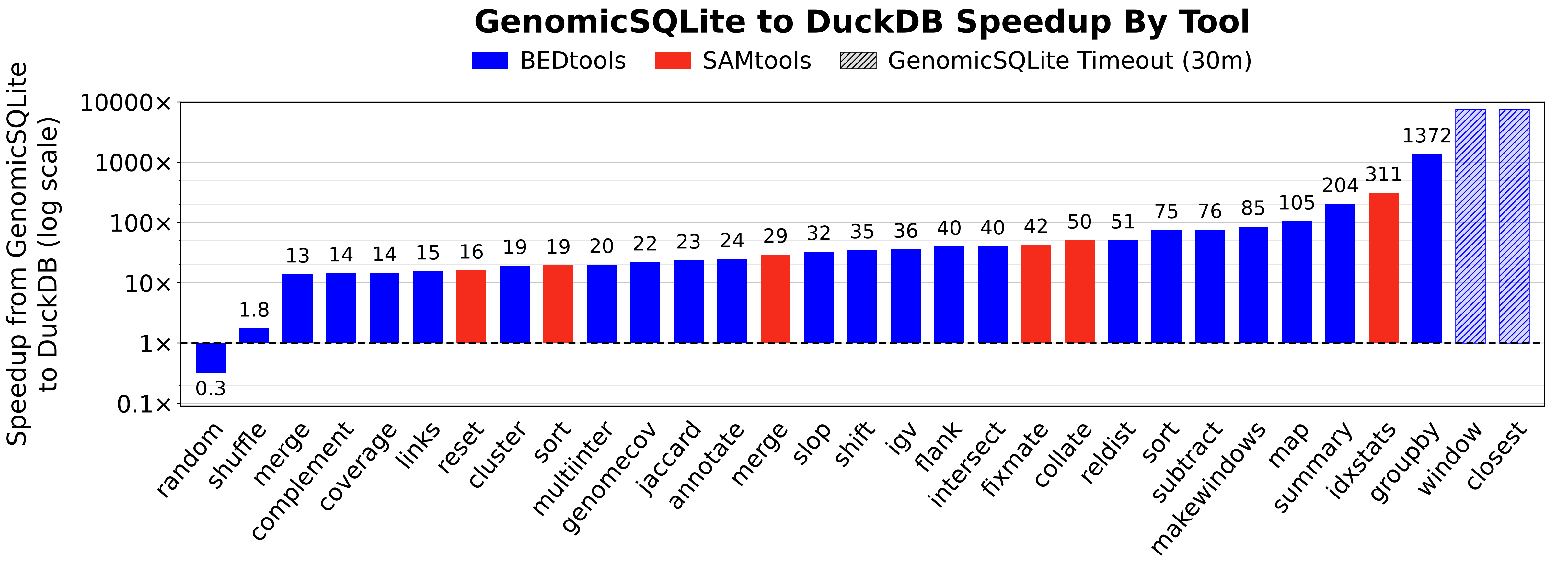}
  \caption{GenomicSQLite to DuckDB speedups by tool.}
  \label{fig:genomicsqlite_to_duckdb_speedups}
\end{figure*}

\begin{figure*}[!t]
  \centering
  \includegraphics[width=0.95\textwidth]{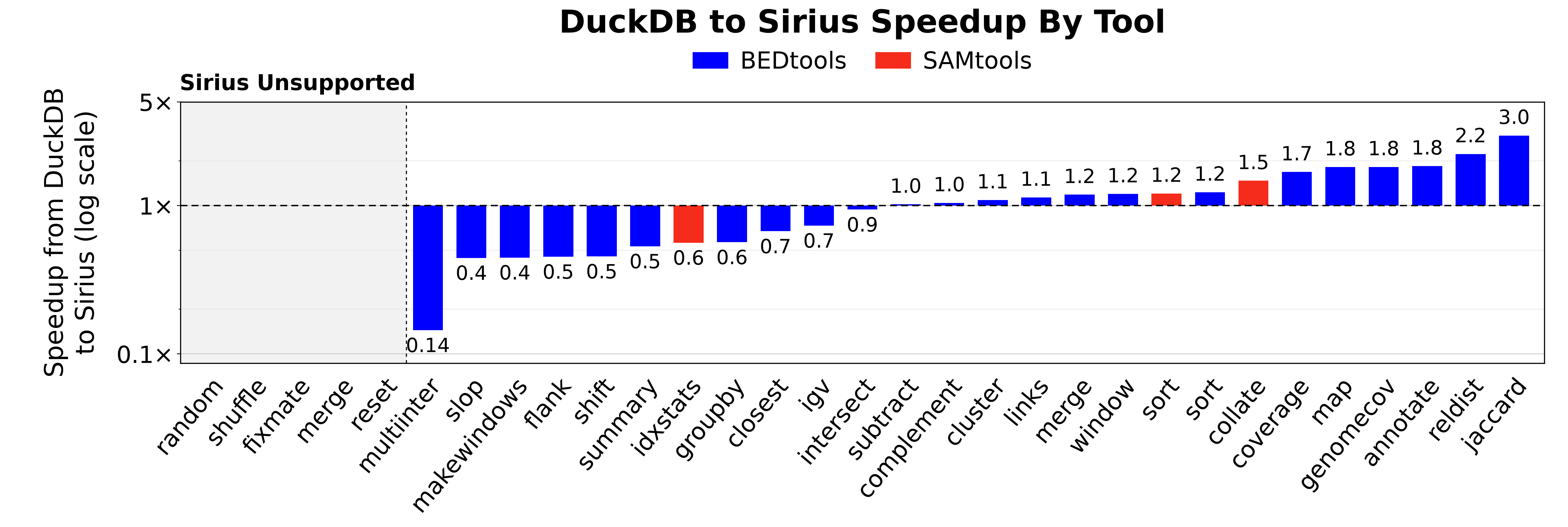}
  \caption{Standard DuckDB to Sirius (GPU) DuckDB speedups by tool. Sirius utilizes RAM-backed spill, sets CPU-fallback to disabled, and pins input data to host and GPU memory where appropriate. Timings do not include input/output.}
  \label{fig:duckdb_to_sirius_speedups}
\end{figure*}

We define the speedup from A to B as the quotient of A's execution time divided by B's execution time. For each CLI operation converted, we calculated the speedups from CLI and GenomicSQLite each to DuckDB as well as the speedup from standard DuckDB to Sirius DuckDB (the latter excluding input/output times as the current Sirius engine does not support the same optimized input/output mechanisms available in standard DuckDB). Fig. \ref{fig:cli_to_duckdb_speedups} plots the CLI to DuckDB speedups, fig. \ref{fig:genomicsqlite_to_duckdb_speedups} plots the GenomicSQLite to DuckDB speedups, and fig. \ref{fig:duckdb_to_sirius_speedups} similarly plots the Standard DuckDB to Sirius speedups: each bar represents the speedup calculation for one operation, with the color (blue or red) representing which tool the operation belongs to (BEDtools or SAMtools) and the y-axis representing the actual speedup magnitude on a logarithmic scale. Bars above the dotted line at y=1 thus indicate a realized speedup, while bars below the dotted line therefore indicate a slowdown. Hatched bars indicate a timeout reached by GenomicSQLite during execution.

In the majority of cases, the DuckDB conversion proved faster than its CLI and GenomicSQLite counterparts (see fig. \ref{fig:cli_to_duckdb_speedups} and \ref{fig:genomicsqlite_to_duckdb_speedups}). Against the original CLI, the operations that simply executed more slowly on DuckDB are SAMtools' fixmate and merge and BEDtools' random, shuffle, and merge. With respect to GenomicSQLite, the only operation that performed worse on DuckDB was BEDtools' random. In turn, outside of input/output times, many of the Sirius GPU-executed queries generally outperformed the standard DuckDB equivalents, with slower performance on some operations being due to optimized standard DuckDB operators that are currently not supported by Sirius (see fig. \ref{fig:duckdb_to_sirius_speedups}). We expect Sirius (and GPU-native databases as a whole) to offer increasingly competitive performance on such queries as they mature and increase support for optimized operations available in the standard versions.

Holistically, the operations that primarily perform variations of scanning and transformations/aggregations experience significant speedups when moved to a relational database engine like DuckDB (as expected). 
However, certain operations involving genomic intervals and multiple datasets perform worse on DuckDB; this is because such operations often require non-equality joins, whose performance depends on the access paths and execution plans selected by the database engine. Our results suggest that existing execution strategies may introduce significant overhead for certain genomic interval workloads, motivating further investigation into genomic-aware query processing and optimization for RDBMS.

\subsection{Experiment 2: Conversion of Example Multi-Variant Workflow}

We implemented the multi-variant workflow discussed previously (see Section II) in BEDtools/SAMtools CLI, GenomicSQLite, DuckDB SQL, and Sirius DuckDB. For these four backends, we benchmarked execution times from start-to-end for five runs after an initial untimed warm-up run. SAMtools utilized seven helper threads for compression (in addition to any concurrency gained by piping processes) while GenomicSQLite utilized the same number of helper threads for compression, sorting, and prefetching purposes. 

For this experiment, we used publicly available ATAC-seq data from ENCODE for two human cell lines, K562 and HepG2, with three biological replicates per cell line. The inputs consisted of six sequencing alignment files (BAM) and their corresponding genomic peak sets (BED), a catalog of population genetic variants with allele-frequency annotations (VCF), and GENCODE gene annotations (GTF). The analysis covered all 22 autosomes of the GRCh38 human reference genome, comprising approximately 2.31 billion alignment records, 979,059 peak intervals, and 132.5 million population variant records. 

\begin{figure}[!t]
  \centering
  \includegraphics[width=\columnwidth]{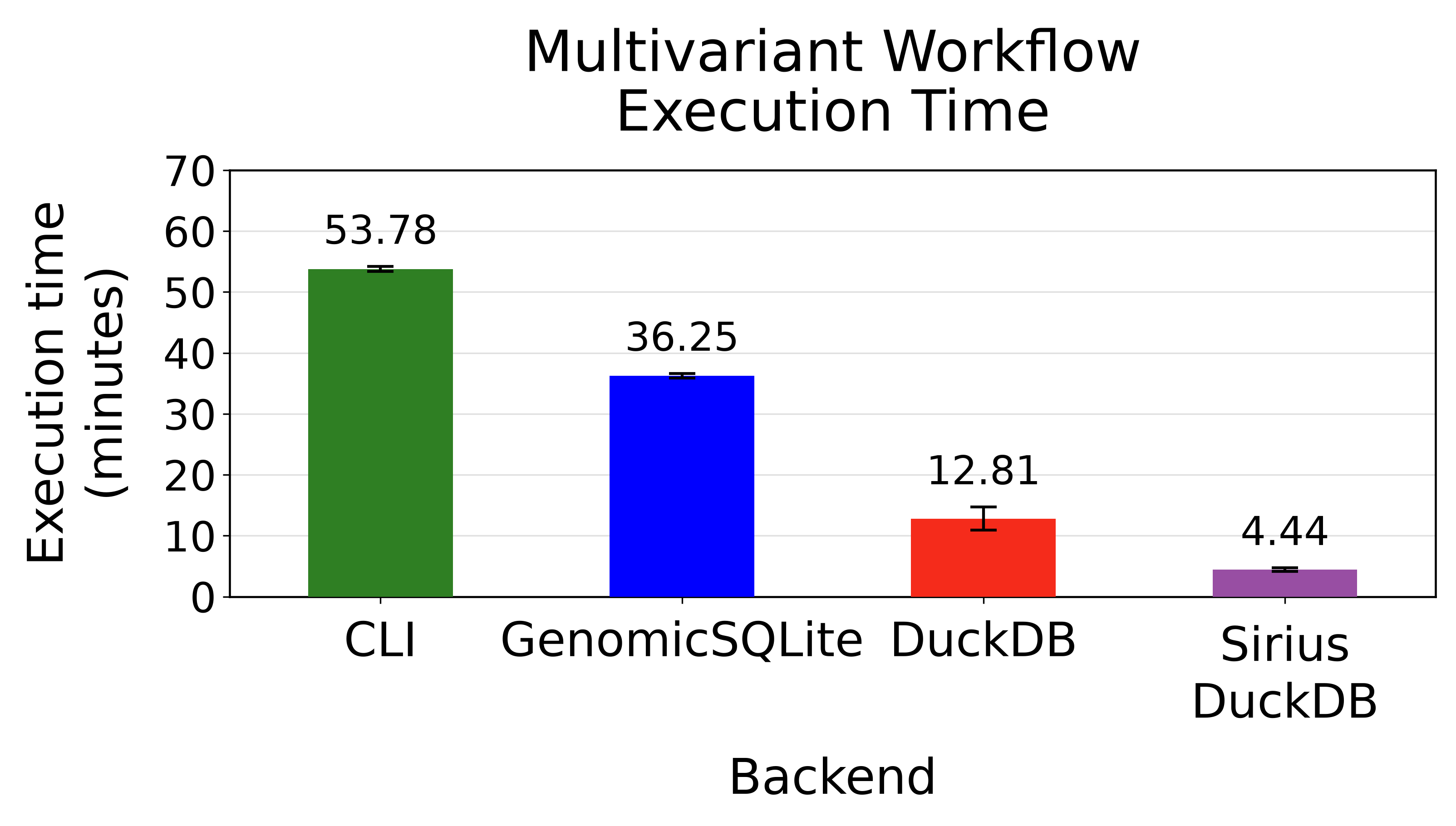}
  \caption{End-to-end execution time of the multi-variant workflow with four different backends. Bars represent mean execution times across five runs, with error bars indicating 95\% confidence intervals. Timings include final output generation but exclude shared input preparation and database loading. Sirius utilizes SSD-backed spill with Parquet files.}
  \label{fig:multi-variant-workflow-timings}
\end{figure}

Fig. \ref{fig:multi-variant-workflow-timings} depicts the average execution time per run for each backend (with 95\% confidence intervals). In particular, all three RDBMS backends vastly outperformed the CLI for this multi-step workflow, with the CLI experiencing significant overhead from repeatedly performing raw file input and output. This evaluation illustrates how off-the-shelf RDBMS and new GPU-native databases can offer increasingly competitive performance on multi-variant, multi-stage workflows in addition to individual operations.

\subsection{Comparison of Composability}
As mentioned previously, many of the tools and binaries utilized as part of the status quo are inflexible, requiring significantly more effort and complexity to emulate functionality not intrinsically provided. This is particularly wasteful when the needed functionality may only be a simple modified variant of an existing command, potentially requiring the creation of a custom executable or pipeline. In contrast, SQL's inherent expressiveness allows for quicker query development, especially in such cases.

For example, the BEDtools 'merge' command by default merges overlapping or adjacent intervals from the same chromosome. Yet, a researcher may only want to merge overlapping or adjacent intervals if they belong to both the same chromosome and the same meta-group (e.g. tissue). Such a modification is a simple addition to the DuckDB query (see listing \ref{lst:composability-sql}, the bolded blue 'tissue' insertions indicate the slight changes that would need to be made to convert 'merge' into the latter version).

\begin{lstlisting}[
    breaklines=true,
    language=SQL,
    float=tbp,
    caption={Example of SQL composability.},
    label={lst:composability-sql},
    emph={tissue},
    emphstyle={\color{blue}\bfseries}
]
WITH bounds AS (
    SELECT *,
        MAX(chromEnd) OVER (
            PARTITION BY chrom, tissue
            ORDER BY chromStart, chromEnd
            ROWS BETWEEN UNBOUNDED PRECEDING
                     AND 1 PRECEDING
        ) AS prev_end
    FROM bed
),
groups AS (
    SELECT *,
        SUM(CASE
            WHEN prev_end IS NULL
              OR chromStart > prev_end
            THEN 1 ELSE 0
        END) OVER (
            PARTITION BY chrom, tissue
            ORDER BY chromStart, chromEnd
            ROWS UNBOUNDED PRECEDING
        ) AS merge_id
    FROM bounds
)
SELECT chrom,
       MIN(chromStart) AS chromStart,
       MAX(chromEnd) AS chromEnd,
       tissue
FROM groups
GROUP BY chrom, tissue, merge_id
ORDER BY chrom, tissue, chromStart;
\end{lstlisting}

If only utilizing the BEDtools binaries, however, perhaps the simplest approach to emulate this functionality would be to combine the chromosome and tissue columns, sort the altered table, run merge, and then split the columns again (see listing \ref{lst:composability-cli}).

\begin{lstlisting}[
    breaklines=true, 
    language=bash,
    float=tbp,
    caption={Example of lack of CLI tools composability.},
    label={lst:composability-cli}
]

awk 'BEGIN {OFS="\t"}
     {print $1 "|" $4, $2, $3}' data.bed \
  | LC_ALL=C sort -k1,1 -k2,2n \
  | bedtools merge -i - \
  | awk 'BEGIN {FS=OFS="\t"}
         {split($1, a, /\|/);
          print a[1], $2, $3, a[2]}' \
  > merged_by_tissue.bed
\end{lstlisting}

\subsection{Comparison of Storage}

\begin{figure}[!t]
  \centering
  \includegraphics[width=\columnwidth]{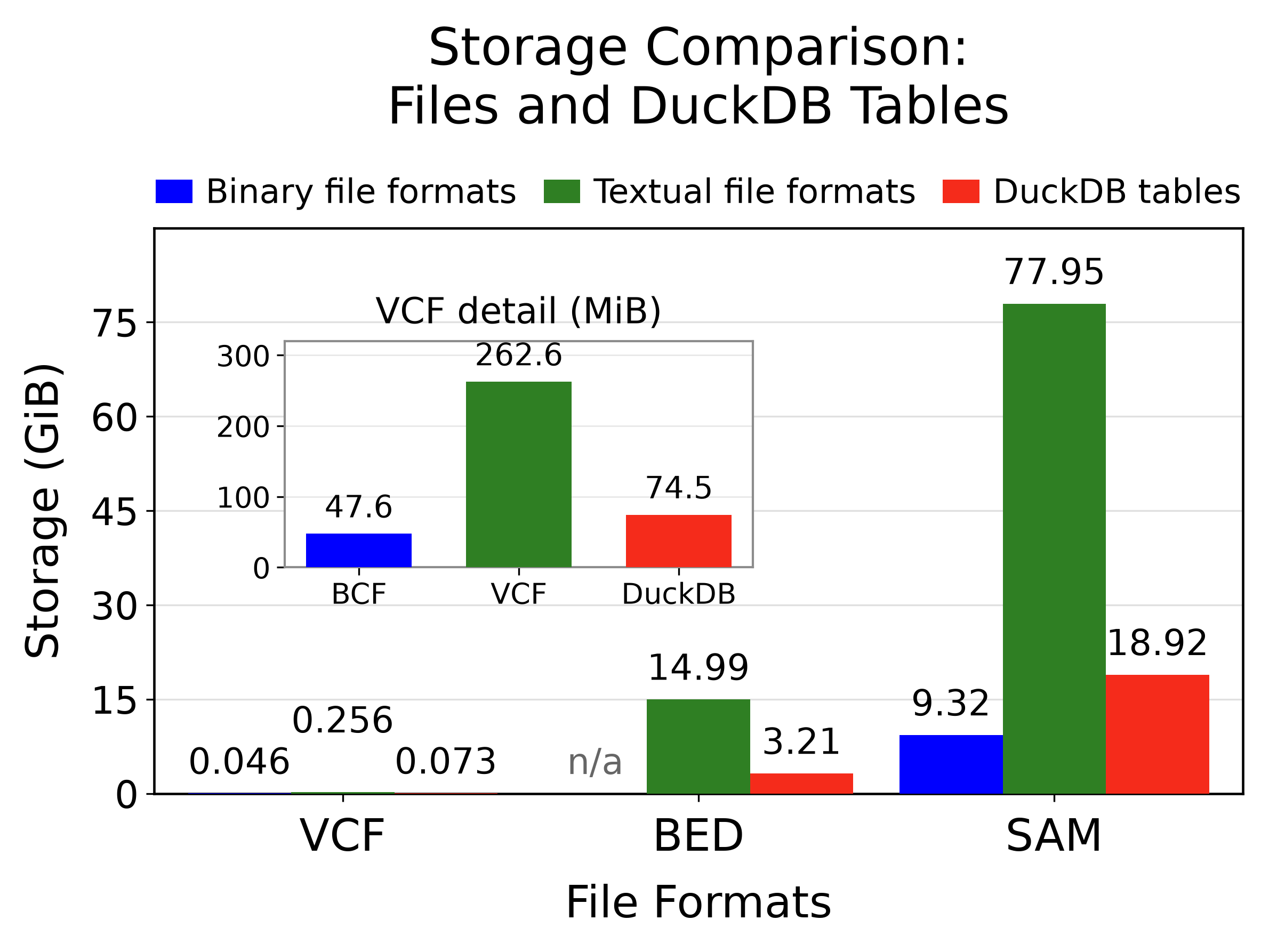}
  \caption{On-disk storage size comparison for equivalent data across three different formats.}
  \label{fig:storage_size_comparison}
\end{figure}

Due to bioinformatics datasets' typically large sizes, there exist specialized compression algorithms that are inadvertently incompatible with many RDBMS. To demonstrate the contrasting storage sizes between the two methods of operation, we measure the on-disk storage costs of binary, textual, and DuckDB-native representations for three genomic data types (VCF, BED, and SAM) where applicable (see fig. \ref{fig:storage_size_comparison}). Through this, we see that, for the same underlying dataset/table, DuckDB may utilize a fraction of the storage space used by the textual format but utilize approximately twice as much storage space as the corresponding binary format.

\section{Future Extensions: Towards Database-Native Genomics}

Our experimental results demonstrate that contemporary, general-purpose relational database management systems can already support a substantial subset of common genomic analysis operations while offering competitive performance and improved expressiveness. However, these experiments also expose several limitations that prevent existing RDBMS from fully replacing specialized genomic analysis tools. In particular, operations involving genomic intervals can require inefficient execution strategies, database-native storage may introduce significant space overhead compared to specialized genomic formats, and certain analytical tasks cannot be naturally expressed through relational operators alone.

We argue that these limitations do not undermine the viability of the relational approach, but instead identify promising opportunities for future database research. In particular, the long-term objective should not simply be to reproduce existing genomic tools using SQL, but to develop a more general data management foundation that combines the expressiveness and optimization capabilities of RDBMS with the specialized requirements of genomic analysis. We outline several research directions toward realizing this vision.

\subsection{Genomic Semantics and Correctness}

As discussed earlier, genomic data introduces semantic challenges that are not automatically resolved by representing information in relational tables. Genomic formats may employ different conventions, schemes, and representations for biological features. For example, BED files conventionally represent intervals using zero-based, half-open coordinates, whereas other formats may employ one-based positions. 

Thus, combining such datasets without accounting for these differences can produce incorrect results even when the corresponding SQL is syntactically valid; a query joining genetic variants with genomic annotations must ensure that both datasets refer to compatible coordinate systems and reference assemblies. Similarly, operations involving strand-specific features, sequence alignments, or overlapping intervals may require semantics that are more specialized than ordinary relational comparisons.

Future research could incorporate genomic semantics directly into logical schemas and database metadata. Rather than relying on application code to interpret coordinate conventions and other biological attributes, RDBMS could explicitly represent these properties and use them during query validation and optimization, with genomic interval types or specialized relational operators potentially providing consistent interpretations of overlap, containment, and distance. 

For example, a database could associate genomic coordinates with explicit coordinate conventions, detecting incompatible joins or automatically rewriting predicates at query planning time. By maintaining explicit representations of source datasets, reference assemblies, and transformations, database systems could additionally make complex analytical workflows easier to inspect, validate, and reproduce.

\subsection{Genomic-Aware Query Processing and Optimization}

Although modern RDBMS have benefited from decades of research in query processing and optimization, their execution engines are primarily designed around general-purpose relational operations. Certain genomic workloads, however, may involve operations that current engines are not designed for.

For example, genomic interval operations (such as the ones in BEDtools) frequently require identifying overlapping or neighboring genomic regions, often across multiple datasets. Such computations often involve complex range predicates and non-equality joins in SQL. While modern RDBMS have developed specialized algorithms for these operations, such as IEJoin \cite{Khayyat2015} in DuckDB, their performance may still vary considerably depending on the underlying data and the execution strategies selected. As demonstrated in our experiments, certain genomic interval operations continue to exhibit substantial performance overheads compared to specialized implementations. 

Future work could investigate how existing relational operators and access methods can be adapted or extended to better support genomic workloads. In particular, chromosome-aware data partitioning, specialized interval indexing structures, and genomic-specific selectivity estimation could improve the efficiency of range-based queries. Moreover, genomic coordinates exhibit dependencies on reference sequences and other biological attributes that may affect data distributions and query selectivity, so incorporating these properties into cost estimation and query planning could allow database engines to more effectively choose between existing general-purpose algorithms and specialized genomic implementations.

This genomic-aware planning would improve the performance of complex genomic workloads while retaining the generality of the relational model. For example, optimal execution for an interval join may depend on the lengths and distribution of intervals within each chromosome, so a genomic-aware optimizer could maintain statistics describing these properties and use them to choose between existing inequality-join algorithms and specialized interval-processing strategies.

\subsection{Genomic Data Representation and Storage}

Existing genomic formats such as BAM, CRAM, and BCF incorporate specialized encoding and compression techniques designed to efficiently represent biological information. While modern RDBMS similarly employ sophisticated storage and compression strategies, these techniques are generally optimized for relational data rather than genomic sequences and alignments.

Our storage experiments illustrate this tradeoff, demonstrating that database-native representations may occupy significantly more space than specialized binary genomic formats (see fig. \ref{fig:storage_size_comparison}). Given the substantial sizes of modern genomic datasets, these storage requirements can represent an important obstacle to widespread RDBMS adoption.

For one, existing database compression techniques should be adapted to exploit genomic-specific properties, potentially incorporating techniques already utilized by established genomic file formats. Alternatively, database engines could provide direct access to existing genomic formats through external tables or specialized scanning operators, allowing users to query data without requiring complete conversion into a database-native representation. This would be most compatible with incremental adoption, as existing genomic datasets would remain in their original formats while exposing their contents through a standardized relational interface. 

However, efficiently executing queries directly against compressed genomic files introduces additional research challenges including predicate pushdown, selective decompression, indexing, and query optimization across heterogeneous storage representations. In particular, efficiency may depend on factors such as query selectivity and the overhead of converting specialized formats into database-native storage. Future work could investigate cost models that account for these factors, allowing the database to determine when direct querying of genomic files is preferable to materializing their contents into relational tables.

\subsection{Interoperability and Hybrid Execution}

Many genomic analysis tasks involve algorithms that cannot be efficiently or conveniently implemented solely in SQL. In these situations, a practical approach involves combining the RDBMS with an external programming language, making use of modern RDBMS mechanisms such as user-defined functions and allowing each paradigm to be utilized where it is most appropriate.

However, simply exposing external functions through SQL does not necessarily provide the full benefits of relational query processing. Instead, future work could investigate allowing database engines to reason about the computational costs, data dependencies, and execution requirements of these external operations. For example, external genomic operators could expose metadata describing their expected input/output schemas or data dependencies. By incorporating this information into query planning, an RDBMS could potentially determine which relational operations should be executed before/after an external operator, reducing unnecessary data transfers or materialization and potentially enabling optimizations across relational and non-relational processing stages.

Interoperability with existing bioinformatics infrastructure is equally important. Established tools and workflow managers such as Snakemake and Nextflow already constitute a substantial ecosystem, and replacing them entirely would introduce considerable adoption costs. Instead, RDBMS should initially serve as complementary components that can be incorporated into existing pipelines, particularly during tertiary analysis. Such an incremental approach would allow researchers to benefit from declarative querying without requiring substantial changes to established computational workflows.
\section{Related Work}

The problems discussed above are well recognized and have received considerable attention from both academia and industry. Existing research has explored various approaches to improving genomic data management, including extending conventional relational database systems, developing specialized declarative query languages, and leveraging alternative storage and execution frameworks. We discuss the most relevant works below and how they relate to our vision of utilizing modern RDBMS as a general-purpose foundation for genomic data analytics.

\subsection{Relational Database Extensions}

Several works have investigated extending existing relational database systems with functionality specifically designed for genomic workloads. These approaches align closely with our vision of utilizing established RDBMS rather than developing entirely new data management platforms.

For example, GenomicSQLite \cite{mlin} extends SQLite with genomic range indexing, specialized compression, and additional functions for common biological operations, allowing genomic datasets to be stored and queried through conventional SQL tables. Similarly, Periscope/SQ \cite{Periscope/SQ} extends PostgreSQL with a specialized query parser and optimizer for sequence datasets. These works further demonstrate the viability of relational databases in genomic analysis as well as the potential for genomic-specific database optimizations. However, while they investigate external functionality for existing databases, we focus on how modern database optimizations can already support common genomic workloads.

\subsection{Domain-Specific Declarative Query Systems}

Other works have instead developed specialized query languages and logical data models to provide higher-level abstractions over genomic data. These approaches offer many of the expressiveness benefits of SQL while incorporating operations and representations specifically designed for biological information.

Genome Query Language (GQL) \cite{GQL}, for example, provides a SQL-like interface for querying genomic alignment and variation data, using logical tables backed by files and metadata. Rather than relying on a conventional database execution engine, GQL translates queries into specialized C++ executables and incorporates algorithms for efficiently processing genomic intervals. SamQL \cite{SamQL} similarly introduces a SQL-like language for SAM/BAM files, allowing complex SAMtools commands and scripts to be expressed declaratively. 

GenoMetric Query Language (GMQL) \cite{GMQL} introduces a more general logical data model for genomic regions and their associated metadata, with specialized operators for querying and combining datasets as part of tertiary genomic analysis. These works demonstrate the advantages of specialized declarative and logical abstractions for bioinformatics, though their use of custom query languages and execution mechanisms introduces an additional layer of complexity for scientists and bioinformaticians.

\subsection{Specialized Storage and Scalable Execution Platforms}

A separate line of research has focused on improving genomic processing through specialized execution algorithms, distributed computing frameworks, and alternative physical storage representations.

Several works incorporate genomic-aware operations into general-purpose analytical engines. GenAp \cite{GenAP}, for example, extends Apache Spark SQL with specialized algorithms for genomic interval joins, significantly improving the performance of range-based queries over conventional execution strategies. Similarly, SeQuiLa \cite{Sequilla} introduces efficient genomic interval processing algorithms within Apache Spark while exposing the operations through an SQL interface, with SeQuiLa-cov \cite{sequilla-cov} building on this approach with optimized depth-of-coverage calculations. These systems demonstrate the performance benefits from combining established analytical engines with domain-specific execution techniques and directly motivate our proposed direction of genomic-aware query optimization within modern RDBMS.

Other Spark-based systems have focused more broadly on supporting scalable genomic workflows. ADAM \cite{ADAM} introduces a logical data model for genomic information backed by Apache Parquet, allowing genomic processing to benefit from columnar storage and distributed execution. Similarly, Glow \cite{Databricks} provides a framework for integrating genomic analysis with Apache Spark, supporting operations over genomic datasets through DataFrame interfaces and integration with external bioinformatics tools. However, these approaches are primarily designed around distributed execution architectures rather than the lightweight, single-node commodity setups that are most frequently utilized by researchers.

Specialized storage systems have also been developed to address the limitations of conventional genomic file formats. TileDB \cite{tiledb.paper}, for example, provides an array-based data management architecture suited for sparse and multidimensional datasets. Its TileDB-VCF library \cite{tiledb_vcf} applies this approach to genomic variant data, enabling compressed storage and efficient retrieval across genomic regions and samples. Commercial analytical platforms such as Snowflake \cite{snowflake} have also been utilized for querying variant datasets in genomic workloads. Additional efforts from industry, such as LatchBio \cite{latchbio} and ExonDB \cite{exondb}, further demonstrate the growing interest in more integrated genomic data management platforms. Together, these approaches illustrate the substantial benefits of specialized storage representations, algorithms, and execution frameworks for genomic analysis, but they also exemplify the sacrifices incurred by tailoring to specific architectures and genomic workloads. 
\section{Conclusion}

In this work, we investigate the viability of utilizing off-the-shelf relational databases for tertiary genomic analysis. By translating common operations from two widely used genomic tools, SAMtools and BEDtools, into SQL and evaluating both individual operations and a multi-stage workflow, we demonstrate that modern RDBMS can provide competitive performance while offering greater expressiveness and composability. We also identify limitations of this approach and highlight research directions further motivating a unified genomic data management platform built upon relational database principles. Just as relational database systems transformed traditional data processing by separating logical queries from physical execution, a similar evolution in genomics could provide a more expressive, extensible, and efficient foundation for future bioinformatics workflows. \\ 

% \clearpage

% =========================================================
% References
% =========================================================

\bibliographystyle{IEEEtran}
\bibliography{sample}

@article{galaxy.orig,
  title={Galaxy: a platform for interactive large-scale genome analysis},
  author={Giardine, Belinda and Riemer, Cathy and Hardison, Ross C and Burhans, Richard and Elnitski, Laura and Shah, Prachi and Zhang, Yi and Blankenberg, Daniel and Albert, Istvan and Taylor, James and others},
  journal={Genome research},
  volume={15},
  number={10},
  pages={1451--1455},
  year={2005},
  publisher={Cold Spring Harbor Lab}
}

@misc{cwl,
    author = {Amstutz, Peter and Crusoe, Michael R. and Tijanić, Nebojša and Chapman, Brad and Chilton, John and Heuer, Michael and Kartashov, Andrey and Kern, John and Leehr, Dan and Ménager, Hervé and Nedeljkovich, Maya and Scales, Matt and Soiland-Reyes, Stian and Stojanovic, Luka},
    title = {Common {Workflow} {Language}, v1.0},
    year = {2016},
    publisher = {Figshare},
    institution = {Common Workflow Language working group},
    doi = {10.6084/m9.figshare.3115156.v2},
    url = {https://doi.org/10.6084/m9.figshare.3115156.v2}
}

@article{nextflow,
  title={Nextflow enables reproducible computational workflows},
  author={Di Tommaso, Paolo and Chatzou, Maria and Floden, Evan W and Barja, Pablo Prieto and Palumbo, Emilio and Notredame, Cedric},
  journal={Nature biotechnology},
  volume={35},
  number={4},
  pages={316--319},
  year={2017},
  publisher={Nature Publishing Group}
}

@article{snakemake,
  title={Sustainable data analysis with Snakemake},
  author={M{\"o}lder, Felix and Jablonski, Kim Philipp and Letcher, Brice and Hall, Michael B and Tomkins-Tinch, Christopher H and Sochat, Vanessa and Forster, Jan and Lee, Soohyun and Twardziok, Sven O and Kanitz, Alexander and others},
  journal={F1000Research},
  volume={10},
  year={2021},
  publisher={Faculty of 1000 Ltd}
}

@article{biopython,
  title={Biopython: freely available Python tools for computational molecular biology and bioinformatics},
  author={Cock, Peter JA and Antao, Tiago and Chang, Jeffrey T and Chapman, Brad A and Cox, Cymon J and Dalke, Andrew and Friedberg, Iddo and Hamelryck, Thomas and Kauff, Frank and Wilczynski, Bartek and others},
  journal={Bioinformatics},
  volume={25},
  number={11},
  pages={1422},
  year={2009},
  publisher={Oxford University Press}
}

@online{sequencing.cost,
    author = {Wetterstrand, Kris A.},
    title = {{DNA Sequencing Costs: Data from the NHGRI Genome Sequencing Program (GSP)}},
    year = {2024},
    url = {www.genome.gov/sequencingcostsdata},
    note = {Accessed December 16, 2024}
}

@online{tiledb_vcf,
    title = {VCF Data Type - TileDB},
    howpublished = {\url{https://tiledb.com/data-types/vcf/}},
    organization = {TileDB, Inc.},
    year = {2024},
    note = {Accessed: December 16, 2024}
}

@unpublished{mlin,
  author    = {Lin, Michea},
  title     = {SQLite: Grand Unified Genomics File Format?},
  note      = {Unpublished poster and a tool.},
url = {https://github.com/mlin/GenomicSQLite},
  year      = {2021}
}

@article{GQL,
    author = {Kozanitis, Christos and Heiberg, Andrew and Varghese, George and Bafna, Vineet},
    title = "{Using Genome Query Language to uncover genetic variation}",
    journal = {Bioinformatics},
    volume = {30},
    number = {1},
    pages = {1-8},
    year = {2013},
    month = {06},
    issn = {1367-4803},
    doi = {10.1093/bioinformatics/btt250},
    url = {https://doi.org/10.1093/bioinformatics/btt250},
    eprint = {https://academic.oup.com/bioinformatics/article-pdf/30/1/1/48913575/bioinformatics\_30\_1\_1.pdf},
}

@article{GMQL,
    author = {Masseroli, Marco and Pinoli, Pietro and Venco, Francesco and Kaitoua, Abdulrahman and Jalili, Vahid and Palluzzi, Fernando and Muller, Heiko and Ceri, Stefano},
    title = "{GenoMetric Query Language: a novel approach to large-scale genomic data management}",
    journal = {Bioinformatics},
    volume = {31},
    number = {12},
    pages = {1881-1888},
    year = {2015},
    month = {02},
    issn = {1367-4803},
    doi = {10.1093/bioinformatics/btv048},
    url = {https://doi.org/10.1093/bioinformatics/btv048},
    eprint = {https://academic.oup.com/bioinformatics/article-pdf/31/12/1881/49013828/bioinformatics\_31\_12\_1881.pdf},
}

@article{Sequilla,
    author = {Wiewiórka, Marek and Leśniewska, Anna and Szmurło, Agnieszka and Stępień, Kacper and Borowiak, Mateusz and Okoniewski, Michał and Gambin, Tomasz},
    title = {SeQuiLa: an elastic, fast and scalable SQL-oriented solution for processing and querying genomic intervals},
    journal = {Bioinformatics},
    volume = {35},
    number = {12},
    pages = {2156-2158},
    year = {2018},
    month = {11},
    issn = {1367-4803},
    doi = {10.1093/bioinformatics/bty940},
    url = {https://doi.org/10.1093/bioinformatics/bty940},
    eprint = {https://academic.oup.com/bioinformatics/article-pdf/35/12/2156/48934835/bioinformatics\_35\_12\_2156.pdf},
}

@article{sequilla-cov,
    author = {Wiewiórka, Marek and Szmurło, Agnieszka and Kuśmirek, Wiktor and Gambin, Tomasz},
    title = {SeQuiLa-cov: A fast and scalable library for depth of coverage calculations},
    journal = {GigaScience},
    volume = {8},
    number = {8},
    pages = {giz094},
    year = {2019},
    month = {08},
    issn = {2047-217X},
    doi = {10.1093/gigascience/giz094},
    url = {https://doi.org/10.1093/gigascience/giz094},
    eprint = {https://academic.oup.com/gigascience/article-pdf/8/8/giz094/29098036/giz094.pdf},
}

@Article{SamQL,
author={Lee, Christopher T.
and Maragkakis, Manolis},
title={SamQL: a structured query language and filtering tool for the SAM/BAM file format},
journal={BMC Bioinformatics},
year={2021},
month={Oct},
day={02},
volume={22},
number={1},
pages={474},
issn={1471-2105},
doi={10.1186/s12859-021-04390-3},
url={https://doi.org/10.1186/s12859-021-04390-3}
}

@article{SAMOriginal,
  author = {Li, H and Handsaker, B and Wysoker, A and Fennell, T and Ruan, J and Homer, N and Marth, G and Abecasis, G and Durbin, R and {1000 Genome Project Data Processing Subgroup}},
  description = {The Sequence Alignment/Map format and SAMtools. - PubMed - NCBI},
  doi = {10.1093/bioinformatics/btp352},
  journal = {Bioinformatics},
  month = aug,
  number = 16,
  pages = {2078-2079},
  pmid = {19505943},
  title = {The Sequence Alignment/Map format and SAMtools},
  url = {https://www.ncbi.nlm.nih.gov/pubmed/19505943},
  volume = 25,
  year = 2009
}

@article{ngs,
  title={Next-generation DNA sequencing techniques},
  author={Ansorge, Wilhelm J},
  journal={New biotechnology},
  volume={25},
  number={4},
  pages={195--203},
  year={2009},
  publisher={Elsevier}
}

@inproceedings{ADAM,
  title={Rethinking Data-Intensive Science Using Scalable Analytics Systems},
  author={Nothaft, Frank A and Massie, Matt and Danford, Timothy and Zhang, Zhao and Laserson, Uri and Yeksigian, Carl and Kottalam, Jey and Ahuja, Arun and Hammerbacher, Jeff and Linderman, Michael and Franklin, Michael and Joseph, Anthony D. and Patterson, David A.},
  booktitle={Proceedings of the 2015 International Conference on Management of Data (SIGMOD '15)},
  year={2015},
  organization={ACM}
}

@article{tiledb.paper,
author = {Papadopoulos, Stavros and Datta, Kushal and Madden, Samuel and Mattson, Timothy},
title = {The TileDB array data storage manager},
year = {2016},
issue_date = {November 2016},
publisher = {VLDB Endowment},
volume = {10},
number = {4},
issn = {2150-8097},
url = {https://doi.org/10.14778/3025111.3025117},
doi = {10.14778/3025111.3025117},
journal = {Proc. VLDB Endow.},
month = nov,
pages = {349–360},
numpages = {12}
}

@online{snowflake,
    author = {Ozer, Stuart},
    title = {Leveraging {Snowflake} to Enable Genomic Analytics at Scale},
    year = {2024},
    howpublished = {\url{https://www.snowflake.com/en/blog/leveraging-snowflake-enable-genomic-analytics-scale/}},
    publisher = {Snowflake Inc.},
    note = {Accessed: December 16, 2024}
}

@online{Databricks,
    author = {Nothaft, Frank Austin and Feng, Karen and Davidge, Henry and Stoica, Ion and Reid, Jeff and Habegger, Lukas and Maxwell, Evan and Barnard, Leland and Kianfar, Kiavash},
    title = {Introducing {Glow}: An Open-Source Toolkit for Large-Scale Genomic Analysis},
    year = {2019},
    month = {10},
    day = {18},
    howpublished = {\url{https://www.databricks.com/blog/2019/10/18/introducing-glow-an-open-source-toolkit-for-large-scale-genomic-analysis.html}},
    publisher = {Databricks},
    note = {Accessed: December 16, 2024}
}

@online{latchbio,
    title = {About {Latch Bio}},
    howpublished = {\url{https://latch.bio/about}},
    organization = {Latch Bio},
    year = {2024},
    note = {Accessed: December 16, 2024}
}

@online{exondb,
    author = {Hauck, Trent},
    title = {Exon Documentation},
    year = {2024},
    howpublished = {\url{https://www.wheretrue.dev/docs/exon/}},
    organization = {WhereTrueSQL},
    note = {Accessed: December 16, 2024}
}

@article{bedtools,
  title={BEDTools: a flexible suite of utilities for comparing genomic features},
  author={Quinlan, Aaron R and Hall, Ira M},
  journal={Bioinformatics},
  volume={26},
  number={6},
  pages={841--842},
  year={2010},
  publisher={Oxford University Press}
}

@Article{GenAP,
author={Kozanitis, Christos
and Patterson, David A.},
title={GenAp: a distributed SQL interface for genomic data},
journal={BMC Bioinformatics},
year={2016},
month={Feb},
day={04},
volume={17},
number={1},
pages={63},
issn={1471-2105},
doi={10.1186/s12859-016-0904-1},
url={https://doi.org/10.1186/s12859-016-0904-1}
}

@Article{BamQL,
author={Masella, Andre P.
and Lalansingh, Christopher M.
and Sivasundaram, Pragash
and Fraser, Michael
and Bristow, Robert G.
and Boutros, Paul C.},
title={BAMQL: a query language for extracting reads from BAM files},
journal={BMC Bioinformatics},
year={2016},
month={Aug},
day={11},
volume={17},
number={1},
pages={305},
issn={1471-2105},
doi={10.1186/s12859-016-1162-y},
url={https://doi.org/10.1186/s12859-016-1162-y}
}

@article{Bioconvert,
    author = {Caro, Hugo and Dollin, Sulyvan and Biton, Anne and Brancotte, Bryan and Desvillechabrol, Dimitri and Dufresne, Yoann and Li, Blaise and Kornobis, Etienne and Lemoine, Frédéric and Maillet, Nicolas and Perrin, Amandine and Traut, Nicolas and Néron, Bertrand and Cokelaer, Thomas},
    title = {BioConvert: a comprehensive format converter for life sciences},
    journal = {NAR Genomics and Bioinformatics},
    volume = {5},
    number = {3},
    pages = {lqad074},
    year = {2023},
    month = {08},
    issn = {2631-9268},
    doi = {10.1093/nargab/lqad074},
    url = {https://doi.org/10.1093/nargab/lqad074},
    eprint = {https://academic.oup.com/nargab/article-pdf/5/3/lqad074/51176799/lqad074.pdf},
}

@misc{ConvertTools,
  author       = {Andrew M. Kropinski},
  title        = {Convert - ONLINE ANALYSIS TOOLS},
  year         = {2025},
  url          = {https://molbiol-tools.ca/Convert.htm},
  note         = {Accessed: 2025-04-14},
}

@inproceedings{Periscope/SQ,
author = {Tata, Sandeep and Lang, Willis and Patel, Jignesh M.},
title = {Periscope/SQ: interactive exploration of biological sequence databases},
year = {2007},
isbn = {9781595936493},
publisher = {VLDB Endowment},
booktitle = {Proceedings of the 33rd International Conference on Very Large Data Bases},
pages = {1406–1409},
numpages = {4},
location = {Vienna, Austria},
series = {VLDB '07}
}

@article{Bonfield2021,
  title = {HTSlib: C library for reading/writing high-throughput sequencing data},
  volume = {10},
  issn = {2047-217X},
  url = {http://dx.doi.org/10.1093/gigascience/giab007},
  doi = {10.1093/gigascience/giab007},
  number = {2},
  journal = {GigaScience},
  publisher = {Oxford University Press (OUP)},
  author = {Bonfield, James K and Marshall, John and Danecek, Petr and Li, Heng and Ohan, Valeriu and Whitwham, Andrew and Keane, Thomas and Davies, Robert M},
  year = {2021},
  month = Jan,
}

@article{Cock2009,
  title = {The Sanger FASTQ file format for sequences with quality scores, and the Solexa/Illumina FASTQ variants},
  volume = {38},
  issn = {1362-4962},
  url = {http://dx.doi.org/10.1093/nar/gkp1137},
  doi = {10.1093/nar/gkp1137},
  number = {6},
  journal = {Nucleic Acids Research},
  publisher = {Oxford University Press (OUP)},
  author = {Cock, Peter J. A. and Fields, Christopher J. and Goto, Naohisa and Heuer, Michael L. and Rice, Peter M.},
  year = {2009},
  month = Dec,
  pages = {1767–1771},
}

@misc{Yogatama2025,
  doi = {10.48550/ARXIV.2508.04701},
  url = {https://arxiv.org/abs/2508.04701},
  author = {Yogatama, Bobbi and Yang, Yifei and Kristensen, Kevin and Sarda, Devesh and Kim, Abigale and Cockcroft, Adrian and Teng, Yu and Patterson, Joshua and Kimball, Gregory and McKinney, Wes and Gong, Weiwei and Yu, Xiangyao},
  title = {Rethinking Analytical Processing in the GPU Era},
  publisher = {arXiv},
  year = {2025},
  copyright = {Creative Commons Attribution 4.0 International},
}

@article{Corces2018,
  title = {The chromatin accessibility landscape of primary human cancers},
  volume = {362},
  issn = {1095-9203},
  url = {http://dx.doi.org/10.1126/science.aav1898},
  doi = {10.1126/science.aav1898},
  number = {6413},
  journal = {Science},
  publisher = {American Association for the Advancement of Science (AAAS)},
  author = {Corces, M. Ryan and Granja, Jeffrey M. and Shams, Shadi and Louie, Bryan H. and Seoane, Jose A. and Zhou, Wanding and Silva, Tiago C. and others},
  year = {2018},
  month = Oct,
}

@article{Fu2014,
  title = {FunSeq2: a framework for prioritizing noncoding regulatory variants in cancer},
  volume = {15},
  issn = {1474-760X},
  url = {http://dx.doi.org/10.1186/s13059-014-0480-5},
  doi = {10.1186/s13059-014-0480-5},
  number = {10},
  journal = {Genome Biology},
  publisher = {Springer Science and Business Media LLC},
  author = {Fu, Yao and Liu, Zhu and Lou, Shaoke and Bedford, Jason and Mu, Xinmeng Jasmine and Yip, Kevin Y and Khurana, Ekta and Gerstein, Mark},
  year = {2014},
  month = Oct,
}

@article{Khayyat2015,
  title = {Lightning fast and space efficient inequality joins},
  volume = {8},
  issn = {2150-8097},
  url = {http://dx.doi.org/10.14778/2831360.2831362},
  doi = {10.14778/2831360.2831362},
  number = {13},
  journal = {Proceedings of the VLDB Endowment},
  publisher = {Association for Computing Machinery (ACM)},
  author = {Khayyat, Zuhair and Lucia, William and Singh, Meghna and Ouzzani, Mourad and Papotti, Paolo and Quiané-Ruiz, Jorge-Arnulfo and Tang, Nan and Kalnis, Panos},
  year = {2015},
  month = Sept,
  pages = {2074–2085},
}

\end{document}